\documentstyle[12pt]{article}
\advance\textheight by 60pt
\advance\voffset by -40pt
\advance\textwidth by 50pt
\advance\oddsidemargin by -25pt
\advance\evensidemargin by -25pt
\def\double_spacesp{\baselineskip 25pt plus 2pt minus 2pt}

\begin{document}
\font\fortssbx=cmssbx10 scaled \magstep2
\hbox to \hsize{
\hskip.5in \raise.1in\hbox{\fortssbx University of Wisconsin - Madison}
\hfill\vbox{\hbox{\bf MAD/PH/744}
            \hbox{\bf SNUTP 93-08}
            \hbox{\bf YUMS 93-03}
            \hbox{June 1993}} }
\vspace{0.5in}
\begin{center}
{\Large {\bf Hadronic $W$ production and the Gottfried Sum Rule}} \\
\vskip 1.0cm
M.~A.~Doncheski$^a$, F.~Halzen$^a$, C.~S.~Kim$^b$ and
M.~L.~Stong$^a$ \\
{\it $^a$Department of Physics, University of Wisconsin, Madison, WI 53706} \\
{\it $^b$Department of Physics, Yonsei University, Seoul 120, Korea}
\end{center}

\vskip 1.0cm
\begin{abstract}

The difference in production rate between $W^+$ and $W^-$ at hadron
colliders is very sensitive to the the difference between up- and
down-quark distributions in the proton.  This sensitivity allows for a
variety of useful measurements.  We consider the difference
$d_s(x,Q^2) - u_s(x,Q^2)$ in the sea distributions and the difference
$\Delta u(x,Q^2) - \Delta d(x,Q^2)$ in the polarized parton distribution
functions.  In both cases we construct an asymmetry to reduce systematic
uncertainties.  Although we discuss measurements at the Tevatron and
future hadron colliders, we find that the Brookhaven Relativistic Heavy
Ion Collider (RHIC) is the most appropriate hadron collider for these
measurements.

\end{abstract}

\newpage
\double_spacesp
\section{Introduction and Results}

It has been appreciated for some time that useful information on the
parton distributions of nucleons can be extracted from measurements of
the charge asymmetry in hadronic $W$
production\cite{berger,proceedings,mrs,mrs92}.  The asymmetry provides us
with information on the difference between the up- and down-quark
structure of the proton, {\it i.e.} on the quantities
$\delta q_v \equiv u_v - d_v$ and $\delta q_s \equiv d_s - u_s$. Here, as
usual, the subscripts $v$ and $s$ refer to the valence and sea components
of the proton structure functions. The quantity $\delta q_v$ is studied
in deep inelastic scattering experiments on proton and neutron targets
and is critical to the determination of the hadronic width of the $W$
from the ratio of the $W$ and $Z$ cross sections in hadron collisions.
While it is often assumed that $\delta q_s = 0$ on the basis of isospin
symmetry, this does not have to be the case.  As $u_v \neq d_v$,
evolution will inevitably result in $u_s \neq d_s$.  Also, as there are
more $u$- than $d$-quarks in the proton, one might imagine that the
further generation of $u\bar{u}$ pairs is suppressed by Fermi
statistics\cite{fieldandfeyn}. As we discuss further on,
experimental evidence for the violation of the Gottfried sum rule has
been interpreted as being the result of the non-vanishing of
$\delta q_s$\cite{mrs,ehq}.

Analogously for $W$ production with polarized beams, a double asymmetry
in the initial proton helicities and in the $W$ charge can provide us
with useful information on the up- and down-quark structure in the
polarized proton.  One of the theoretical assumptions generally used in
the interpretation on the EMC result on proton spin is based on a
combination of low energy experiment and isospin symmetry, {\it i.e.} the
difference between the polarized $u$- and $d$ quark distributions.  This
quantity can be directly measured at higher energy in $W$ production
at a polarized proton-proton collider.

The main point of this paper is that the $W$ charge asymmetry in hadron
collisions measures both $\delta q_v$ and $\delta q_s$.  We show,
moreover, that the latter quantity can only be effectively probed {\bf in
proton-proton colliders}.  $W$ production at RHIC thus has unique
aspects.  This includes the possibility of using polarized beams, which
we discuss in some detail.

\section{Gottfried sum rule in unpolarized production}

The recent NMC\cite{nmc} result on the Gottfried\cite{gottfried} sum rule
\begin{equation}
I_G(x) = \frac{1}{3} \int_x^1 dy [u_v(y) - d_v(y)]
        + \frac{2}{3} \int_x^1 dy [\bar{u}(y) - \bar{d}(y)]
\end{equation}
(extrapolated to $x=0$) of $I_G(0) = 0.240 \pm 0.016$ is significantly
different from the result of 1/3 expected if the usual assumptions are
made, {\it i.e.} the proton consists of valence $uud$ and
$\bar{u}(y) = \bar{d}(y)$.  This ``violation'' of the Gottfried sum rule
is quite easy to understand --- there are more $u$-quarks in the proton
than $d$-quarks, so the Pauli exclusion principle suppresses the
production of $u \bar{u}$ pairs in the sea relative to $d \bar{d}$
pairs\cite{fieldandfeyn} --- but an independent confirmation of this
result is desirable.

The large number of $W$'s (we estimate 3000 for $\sqrt{s} = 200$~GeV and
300,000 for $\sqrt{s} = 500$~GeV) to be produced at the Brookhaven
Relativistic Heavy Ion Collider (RHIC) will provide the necessary
confirmation as has been suggested in the literature\cite{bs}.  Instead
of considering the separate production of $W^+$ and $W^-$, consider the
charge asymmetry $A_W$\cite{berger}:
\begin{eqnarray}
N_W & = & \frac{d \sigma(W^+)}{dy_W} - \frac{d \sigma(W^-)}{dy_W}
\nonumber \\
D_W & = & \frac{d \sigma(W^+)}{dy_W} + \frac{d \sigma(W^-)}{dy_W}
\nonumber \\
A_W & = & \frac{N_W}{D_W}.
\end{eqnarray}
It is straightforward to show that (to simplify the notation we suppress
the dependence of the parton distribution functions on $Q^2$ in the
following):
\begin{eqnarray}
N_W & \propto & \cos^2 \theta_c \{u_s(x_2) \; [u_v(x_1) - d_v(x_1)] \;+\;
    u_v(x_2) \; [d_s(x_1) - u_s(x_1)] \} \nonumber \\
    & + & \sin^2 \theta_c \{u_v(x_1) s_s(x_2) \; - \; c_s(x_1) d_v(x_2)
    \} \nonumber \\
    & + & (x_1 \leftrightarrow x_2)
\end{eqnarray}
and
\begin{eqnarray}
D_W & \propto & \cos^2 \theta_c \{u_s(x_2) \; [u_v(x_1) + d_v(x_1)] \;+\;
    u_v(x_2) \; [d_s(x_1) + u_s(x_1)] \; + \; 2 s_s(x_1) c_s(x_2) \}
    \nonumber \\
    & + & \sin^2 \theta_c \{u(x_1) s_s(x_2) \; + \; c_s(x_1) d(x_2) \}
    \nonumber \\
    & + & (x_1 \leftrightarrow x_2),
\end{eqnarray}
where $\theta_c$ is the Cabibo angle.  This gives the exact expression we
use to calculate $A_W$.  In calculating these cross sections we include
the usual $K$-factor for $W$ production,
$K = 1 + 8 \pi \alpha_s(M_W^2)/9$ (see, {\it e.g.}, Ref.~\cite{berger}).

To further illustrate our analysis, consider only the terms proportional
to $\cos^2 \theta_c$.  This is a relatively good approximation in any
case as $\sin^2 \theta_c \sim 0.05$.  In this limit, the numerator factor
reduces to
\begin{equation}
N_W \; \approx \; u_s(x_2) \delta q_v(x_1) \; + \; u_v(x_2) \delta
q_s(x_1) \; + \; (x_1 \leftrightarrow x_2)
\end{equation}
where $\delta q_v(x) = u_v(x) - d_v(x)$ and
$\delta q_s(x) = d_s(x) - u_s(x)$.  $W$ production is central and
relatively flat in rapidity, so we examine $A_W$ at $y=0$.  Here, $x_1$ =
$x_2$ = $x_0 \equiv \frac{\mbox{$M_W$}}{\mbox{$\sqrt{s}$}}$, and
\begin{equation}
A_W \mid_{y=0} \; \approx \; \frac{u_s(x_0) \delta q_v(x_0) \; + \;
u_v(x_0) \delta q_s(x_0)}
{u_s(x_0) \; [u_v(x_0) + d_v(x_0)] \; + \; u_v(x_0) \;
[d_s(x_0) + u_s(x_0)]}.
\end{equation}

A similar asymmetry can be formed at $p \bar{p}$ colliders.  In this
case, the numerator factor is
\begin{eqnarray}
N_W & \propto & \cos^2 \theta_c \{ u_s(x_2) \delta q_v(x_1)
      \; + \; \frac{1}{2} \delta q_v(x_1) \; (u_v(x_2) + d_v(x_2))
      \; + \; u_v(x_1) \delta q_s(x_2) \} \nonumber \\
    & + & \sin^2 \theta_c \{ u_v(x_1) s_s(x_2) \; + \; c_s(x_1) d_v(x_2)
    \} \nonumber \\
    & - & (x_1 \leftrightarrow x_2).
\end{eqnarray}
The expression separates into terms proportional to $\delta q_v$ and
$\delta q_s$ and another small term proportional to $\sin^2 \theta_c$,
but it is not as simple as the expression (5) for the asymmetry at $pp$
colliders.  More importantly, the asymmetry vanishes for $y=0$ at
$p \bar{p}$ colliders.  In the central region of rapidity, where most of
the $W$'s are produced, the asymmetry is small, severely limiting the
usefulness of $p \bar{p}$ colliders in this measurement.

We have demonstrated the usefulness of $pp$ colliders in the measurement
of the difference in the distributions $u_s$ and $d_s$.  RHIC has been
designed with a variable center-of-mass energy, ranging from 50-500~GeV.
This gives an $x_0$ range $0.16 \leq x_0 \leq 1$ (though with low
statistics at threshhold).  SSC and LHC are sensitive to $x_0 = 0.002$
and $0.005$ respectively (although moving away from $y=0$ allows for a
range of sensitivity about these values).  At such small values of $x$,
sea quarks dominate $W$ production.  Referring to Eqn. (6), it is clear
that the term $u_s(x_0) \delta q_v(x_0)$ will easily dominate the
asymmetry, leading to very little sensitivity to $\delta q_s(x_0)$ at SSC
and LHC.  This argument does not hold at RHIC since the larger $x$ region
probed leads to a larger valence contribution, which in turn leads to a
rather large contribution from the difference in the light sea quark
distributions.  Thus RHIC is the only $pp$ collider where this
measurement can be performed.

Our conclusion thus far is that RHIC is uniquely suited for this
measurement, because a) $A_W$ is large at small rapidity where the
statistics are best, unlike all $p \bar{p}$ colliders, b) RHIC probes the
correct $x$-range, unlike $pp$ supercolliders and c) RHIC has variable
center-of-mass energy, which allows for measurement at different $x$
without measuring at different rapidity.

\section{Results on the Gottfried Sum Rule}

Consider the asymmetry $A_W$ at the Tevatron.  In Fig.~1 we show the
cross section for $W^+$ and $W^-$ production, for various parton
distribution functions.  Next, we show the numerator factor $N_W$ and the
dominant contribution from $\delta q_s$ (Fig.~2).  The contribution from
a non-$SU(2)$ symmetric sea is very small compared to the contribution
from the difference in valence distributions, $\delta q_v$.  As we need
to measure this asymmetry at large rapidity ($\sim 1.5$), we see that
information on $d_s - u_s$ will be difficult to obtain from the Tevatron.

Some data exist\cite{cdf} on the leptonic asymmetry $A_{\ell}$ at the
Tevatron.  Here,
\begin{eqnarray}
N_\ell & = & \frac{\mbox{$d \sigma(\ell^-)$}}{\mbox{$dy_\ell$}}
          - \frac{\mbox{$d \sigma(\ell^+)$}}{\mbox{$dy_\ell$}} \nonumber
          \\
D_\ell & = & \frac{\mbox{$d \sigma(\ell^-)$}}{\mbox{$dy_\ell$}}
          + \frac{\mbox{$d \sigma(\ell^+)$}}{\mbox{$dy_\ell$}} \nonumber
          \\
A_\ell & = & \frac{N_\ell}{D_\ell}
\end{eqnarray}
is the charge asymmetry in rapidity distribution for single lepton
production.  We show in Fig.~3 our result for $A_e$, the asymmetry for
electrons and compare it to the data from CDF\cite{cdf}.  In Fig.~4 we
compare the contribution to $N_e$ from $\delta q_s$ with the total
numerator.

Next, consider the asymmetry $A_W$ at hadron supercolliders.  Our results
for the SSC are also representative of the results that can be obtained
at the LHC.  In Fig.~5 we show the numerator factor $N_W$ and the
dominant contribution from $\delta q_s$.  It is clear that the
contribution from a non-$SU(2)$ symmetric sea is again very small
compared to the contribution from the difference in valence
distributions, $\delta q_v$.  Although the statistics will be very good,
the relative smallness of $\delta q_s$ will make this measurement
difficult.

Finally, we consider the asymmetry $A_W$ at RHIC, a high luminosity
(${\cal L} = 2 \times 10^{32} \; {\rm cm}^{-2} {\rm sec}^{-1} = 6000 \;
{\rm pb}^{-1}/{\rm yr}$) collider capable of producing proton-proton
collisions for center-of-mass energies between 50 and 500~GeV.  RHIC will
also collide protons on heavy nuclei in order to connect to existing data
and heavy ions on heavy ions.  We assume a nominal running time of two
months at full luminosity.  In order to be somewhat conservative, we
estimate event numbers based on 300~pb$^{-1}$ integrated luminosity.  We
assume a generic collider-type detector, and we require the photons and
electrons observed to lie in the rapidity range $|y| \leq 2$.  This
simulates the acceptance of the proposed STAR detector at RHIC, level 2
for photons and electrons\cite{aki}.  We do not consider the possibility
of the detection of muons at RHIC.  We show the numerator factor $N_W$,
and the contribution to $N_W$ from $\delta q_s$ in Figs.~6a and 6b
respectively for $\sqrt{s} = 200$~GeV and 500~GeV.  At RHIC, the
contributions from $\delta q_s$ are in no way suppressed; they can in
fact dominate $N_W$.  The parton distributions with a symmetric sea
({\it e.g.}, HMRSB, EHLQ2 and MRSD0) generally give smaller asymmetries
(especially at small $W$ rapidity, $y_W$) than the distributions with
non-symmetric sea ({\it e.g.}, MRSDM, CTEQ1L and CTEQ1M).  Fig.~7 gives
the asymmetry $A_W$ for $\sqrt{s} = 500$~GeV for various choices of the
parton distribution functions.  Here at last we can observe in the
asymmetry the effects of the non-symmetric sea, although for this
particular scenario, the $\delta q_s$ contribution using the CTEQ1L
distributions accidentally small.  The explanation for this phenomenon
is given in Ref.~\cite{cteq} - the sign of $\bar{d} - \bar{u}$ changes
with $x$, and $W$ production at $\sqrt{s} = 500$~GeV is sensitive to
values of $x$ near the crossover point for the CTEQ1L distributions.

Having determined that this measurement is in principle possible at RHIC,
we study events with a single high-$p_{_T}$ lepton and missing $p_{_T}$.
In Figs.~8a and 8b we show the rapidity distributions for production of
$e^+$ and $e^-$ at center of mass energies of 200 and 500~GeV,
respectively.  Even in the worst case ($e^-$ production at 200~GeV), we
expect about 15 events in the rapidity range $|y_e| < 0.5$, so the
statistics should be sufficient.  In Figs.~9 we compare the
contributions to $N_\ell$.  Again, the parton distributions with a
symmetric sea give generally smaller asymmetries that the parton
distributions with non-symmetric sea, as can be seen from the
$\delta q_s$ contributions in Fig.~9, although the accidentally small
$\delta q_s$ contribution from CTEQ1L distributions at
$\sqrt{s} = 500$~GeV is apparent.  Finally, in Fig.~10 we show the
observable asymmetry $A_e$ for center-of-mass energies of 200 and
500~GeV.  Here, especially at small electron rapidity $y_e$, the
asymmetries from the symmetric sea parton distributions cluster
together and are reasonably separated from the asymmetries from the
non-symmetric sea parton distributions.

\section{$\Delta u - \Delta d$ in polarized production}

The interpretation of the EMC result on proton spin relies upon two low
energy quantities, namely $(\Delta u - \Delta d) = g_A$ (from isospin
invariance) and $(\Delta u + \Delta d - 2 \Delta s) = 3 \; F \;-\; D$
(from $SU(3)$ symmetry) where $g_A$ is the axial-vector coupling in
neutron $\beta$-decay and $F$ and $D$ are the invariant amplitudes for
the axial-vector current in hyperon semileptonic decays.  Here, in the
usual notation
\begin{equation}
\Delta q(\mu^2) s_{\alpha} = \langle p,s \mid \bar{q} \gamma_\alpha
\gamma_5 q \; \mid p,s \rangle \mid_{\mu^2},
\end{equation}
where $\mid p,s \rangle$ is a proton state with spin vector $s_\alpha$.
The quantity $\Delta q$ is related to the polarized parton distribution
function for quark $q$
\begin{equation}
\Delta q(Q^2) = \int_0^1 \Delta q(x,Q^2) dx
\end{equation}
where $\Delta q(x,Q^2)$ is the difference in the distributions of a quark
$q$ in a longitudinally polarized proton with the same and opposite
helicity as the proton.  As long as the assumptions of isospin invariance
and $SU(3)$ symmetry are valid, these particular (non-singlet) combinations
of the $\Delta q$'s will not run with $Q^2$; they reduce to the
difference and sum of the first moments of the polarized up- and
down-valence distributions.  Thus, the two quantities above can be used
at higher energy even thought they are extracted from low energy data.
It would, nonetheless, be comforting to extract one (or both) of these
quantities at a higher energy as a consistency check.

Analogous to the earlier discussion of unpolarized $W$ production we can
derive the following for polarized $W$ production:
\begin{eqnarray}
\frac{d \sigma^{++}(W^+)}{dy_W} - \frac{d \sigma^{+-}(W^+)}{dy_W}
& \propto & \cos^2 \theta_c \Delta u(x_1) \Delta \bar q(x_2)
   \; + \; \sin^2 \theta_c \Delta u(x_1) \Delta s(x_2)
   \; + \; (x_1 \leftrightarrow x_2) \nonumber \\
\\
\frac{d \sigma^{++}(W^-)}{dy_W} - \frac{d \sigma^{+-}(W^-)}{dy_W}
& \propto & \cos^2 \theta_c \Delta d(x_1) \Delta \bar q(x_2)
   \; + \; \sin^2 \theta_c \Delta \bar q(x_1) \Delta s(x_2)
   \; + \; (x_1 \leftrightarrow x_2). \nonumber \\
\end{eqnarray}
We assume that the charm content of the proton is zero, which is a very
good approximation below Tevatron energies, and $\Delta \bar{u}
= \Delta \bar{d} = \Delta u_s = \Delta d_s \equiv \Delta \bar q$, and we
again suppress the dependence of the parton distriubtion functions on
$Q^2$.  In the usual notation, $\sigma^{++}$ ($\sigma^{+-}$) is the cross
section for two protons with the same (opposite) helicities and the
polarized parton distribution functions as described above.  As before,
we include the standard $K$-factor in our calculation of the cross
section.  Just as the difference of cross sections for unpolarized $W$
production is the most interesting and useful quantity, we find that the
difference
\begin{eqnarray}
& & \left\{ \left[ \frac{d \sigma^{++}(W^+)}{dy_W} -
        \frac{d \sigma^{+-}(W^+)}{dy_W} \right]
- \left[ \frac{d \sigma^{++}(W^-)}{dy_W} -
      \frac{d \sigma^{+-}(W^-)}{dy_W} \right] \right\} \propto
\nonumber \\
\nonumber \\
& & \{ \; \cos^2 \theta_c \; (\Delta u(x_1) - \Delta d(x_1) ) \;
      \Delta \bar q(x_2) \; + \; \sin^2 \theta_c \; \Delta u_v(x_1) \;
      \Delta s(x_2) \; + \; (x_1 \leftrightarrow x_2) \}
\end{eqnarray}
is particularly useful in isolating the differences we would like to
measure.

Consider the quantity ${\cal A}_{LL}$:
\begin{equation}
{\cal A}_{LL} = \left\{ \frac{ \left[
        \frac{\mbox{$d \sigma^{++}(W^+)$}}{\mbox{$dy_W$}} -
        \frac{\mbox{$d \sigma^{+-}(W^+)$}}{\mbox{$dy_W$}} \right]
        - \left[ \frac{\mbox{$d \sigma^{++}(W^-)$}}{\mbox{$dy_W$}} -
         \frac{\mbox{$d \sigma^{+-}(W^-)$}}{\mbox{$dy_W$}} \right]}
         {\frac{\mbox{$d \sigma^{++}(W^+)$}}{\mbox{$dy_W$}} +
         \frac{\mbox{$d \sigma^{+-}(W^+)$}}{\mbox{$dy_W$}}
         + \frac{\mbox{$d \sigma^{++}(W^-)$}}{\mbox{$dy_W$}} +
          \frac{\mbox{$d \sigma^{+-}(W^-)$}}{\mbox{$dy_W$}} } \right\}.
\end{equation}
This asymmetry is again dominated by the $\cos^2 \theta_c$ terms, not
only because $\sin^2 \theta_c$ is small but also because
$\Delta u(x)$ and $\Delta d(x)$ have opposite signs, so the
interference is constructive.  In the limit that we only consider the
$\cos^2 \theta_c$ terms, this asymmetry reduces to
\begin{equation}
{\cal A}_{LL} \; \approx \; \left\{
\frac{ [ \Delta u(x_1) - \Delta d(x_1) ] \Delta q(x_2) +
(x_1 \leftrightarrow x_2) }
{ [ u(x_1) + d(x_1) ] q(x_2) +
(x_1 \leftrightarrow x_2) } \right\}.
\end{equation}
Finally, because the cross section for central $W$'s is large, we can
examine this asymmetry at $y=0$, which gives:
\begin{equation}
{\cal A}_{LL} \mid_{y=0} \; \approx \;
\frac{ [ \Delta u(x_0) - \Delta d(x_0) ] \Delta q(x_0)}
{ [ u(x_0) + d(x_0) ] q(x_0)},
\end{equation}
where $x_0 = \frac{\mbox{$M_W$}}{\mbox{$\sqrt{s}$}}$ as before.
Inclusion of the $\sin^2 \theta_c$ terms will slightly modify this
result, but it should be possible to extract the combination
$\Delta u(x) - \Delta d(x)$ from the data due to the large statistics.
Furthermore, by running RHIC at several center-of-mass energies and
moving away from $y=0$, it will be possible to map the desired polarized
parton distributions for a range of $x$ values.  We give the expressions
above under the approximation $\sin^2 \theta_c = 0$ merely to simplify
the discussion.  In the figures we use the exact expressions.

\section{Results on $\Delta u - \Delta d$}

A program of polarized proton-proton collisions, at full energy and
luminosity, is being discussed\cite{rsc}.  We show in Fig.~11, the $W$
rapidity distribution, $d \sigma/dy_W$, for $W^+$ and $W^-$ production at
two representative RHIC energies, 500~GeV and 200~GeV.  Given the large
integrated luminosity from a two month run at RHIC
($\sim 300 \; pb^{-1}$), there will be a rather large number of
$W$'s produced at RHIC.  Next, we present our results for ${\cal A}_{LL}$
as a function of the $W$ rapidity $y_W$ for RHIC at $\sqrt{s} = 200$~ GeV
(Fig.~12a) and at $\sqrt{s} = 500$~GeV (Fig.~12b).  In Fig.~12, we
use two extreme cases for the polarized parton distribution functions,
one in which the polarized gluon is taken to be large (and consequently
the polarized strange quark distribution is small - set $I_S$=0 of
Bourrely, Guillet and Chiappetta ({\em BGC})\cite{bgc}) and one in which
the polarized gluon is small (and the polarized strange quark
distribution large - set $I_S$=1 of BGC).  We note that the large
difference in the asymmetries calculated using the different polarized
parton distribution functions is due primarily to the large uncertainty
in the polarized sea distribution.  This uncertainty can be traced back
to the different explanations of the EMC effect on proton spin.  This
uncertainty will be reduced using other measurements at RHIC, {\it e.g.},
jet\cite{kunszt,guillet,sofferrhic}, direct
photon\cite{sofferrhic,qiu,was}, heavy quark\cite{contogouris} or
charmonium production\cite{cortes,Russians,doncheski,us}, designed to
measure the polarized gluon contribution to proton spin.  Once this error
is reduced, it will be possible to extract the desired quantity,
$\Delta u - \Delta d$.  Also, the polarized parton distribution set with
the `smaller' polarized sea quark gives a larger asymmetry at large
rapidity for $\sqrt{s} = 200$~GeV.  This is a consequence of the
different large-$x$ behavior in the polarized parton distributions; the
event rate at large rapidity is sufficiently small that for practical
purposes, this region of large rapidity will contribute very little to
the observed asymmetry.

Of course, the experiments at RHIC will not directly observe $W$'s, but
rather will reconstruct them from the sample of single isolated lepton
and missing $p_{_T}$, so one should also study the possibilities
including the decay.  To simulate the aceptance of the STAR detector, we
impose a cut of $|y| \leq 2$ on electrons; STAR will have no acceptance
for muons\cite{aki}.  In addition, we place a cut on the transverse mass
of the reconstructed $W$ of 50~GeV, which will remove much of the
background without significantly decreasing the signal.  In Fig.~13 we
show the rapidity distribution, $d\sigma/dy_e$, for single electrons and
positrons for two representative center of mass energies, 500 and
200~GeV.  Even including the $W$ decay there will still be a large number
of events (about 130 for $\sqrt{s} = 200$~GeV and 13000 for
$\sqrt{s} = 500$~GeV).  We give, in Fig.~14 the asymmetry
${\cal A}^e_{LL}$ including electronic decay of the $W$, with
$\sqrt{s} = 200$~GeV (Fig.~14a) and with $\sqrt{s} = 500$~GeV
(Fig.~14b).  The curves follow the convention of Fig.~12.  Here,
\begin{equation}
{\cal A}^e_{LL} = \left\{ \frac{ \left[
        \frac{\mbox{$d \sigma^{++}(e^+)$}}{\mbox{$dy_e$}} -
        \frac{\mbox{$d \sigma^{+-}(e^+)$}}{\mbox{$dy_e$}} \right]
        - \left[ \frac{\mbox{$d \sigma^{++}(e^-)$}}{\mbox{$dy_e$}} -
         \frac{\mbox{$d \sigma^{+-}(e^-)$}}{\mbox{$dy_e$}} \right]}
         {\frac{\mbox{$d \sigma^{++}(e^+)$}}{\mbox{$dy_e$}} +
         \frac{\mbox{$d \sigma^{+-}(e^+)$}}{\mbox{$dy_e$}}
         + \frac{\mbox{$d \sigma^{++}(e^-)$}}{\mbox{$dy_e$}} +
          \frac{\mbox{$d \sigma^{+-}(e^-)$}}{\mbox{$dy_e$}} } \right\}.
\end{equation}
Again, the polarized parton distributions that give use a large polarized
gluon to explain the EMC effect on the spin of the proton give give a
small asymmetry due to the presence of a $\Delta \bar{q}$ factor in the
asymmetry, and at $\sqrt{s} = 200$~GeV the different large-$x$ behavior
of the polarized parton distributions is apparent in the crossings of
the asymmetries.  Once the uncertainties in the polarized gluon and sea
distributions are reduced, it will be possible to extract
$\Delta u - \Delta d$.

\begin{center}
{\Large {\bf Acknowledgements}}
\end{center}

The work of MAD, FH and MLS was supported in part by the U.~S. Department
of Energy under Contract No.~DE-AC02-76ER00881, in part by the Texas
National Research Laboratory Commission under Grant Nos.~RGFY9173,
RGFY9273 and RGFY93-221, and in part by the University of Wisconsin
Research Committee with funds granted by the Wisconsin Alumni Research
Foundation.  The work of CSK was supported in part by the Korean Science
and Engineering Foundation, in part by the Korean Ministry of Education,
in part by the Center of Theoretical Physics at Seoul National University
and in part by a Yonsei University Faculty Research Grant.

\newpage

\begin{center}
{\Large {\bf Figure Captions}}
\end{center}

\begin{itemize}

\item[Figure 1]{Rapidity distribution for the production of $W^-$ and
$W^+$ at the Tevatron.  The different curves correspond to different
parton distribution functions: dotted - HMRSB\cite{hmrs}, double dotted -
EHLQ set 2\cite{ehlq}, solid - MRSD0\cite{mrs92}, dashed -
MRSDM\cite{mrs92}, dot-dashed - CTEQ1L\cite{cteq} and double
dot-dashed - CTEQ1M\cite{cteq}.}

\item[Figure 2]{Numerator factor $N_W$ (as defined in the text) versus
$y_W$ at the Tevatron.  The different curves follow the convention of
Figure~1.  Both the total $N_W$ and the contribution from $\delta q_s$
are shown.}

\item[Figure 3]{Electron asymmetry $A_e$ at the Tevatron.  The labeling
of the different curves follows the convention of Figure~1.  Data are
from Ref.~\cite{cdf}.}

\item[Figure 4]{Numerator factor $N_e$ (as defined in the text) versus
$y_e$ at the Tevatron.  The different curves follow the convention of
Figure~1.  Both the total numerator factor and the contribution from
$\delta q_s$ are shown.}

\item[Figure 5]{As Fig.~2, at SSC.}

\item[Figure 6]{As Figs.~2,~5, for a) RHIC at 200~GeV and b) RHIC at
500~GeV.}

\item[Figure 7]{The charge asymmetry $A_W$ (as defined in the text)
versus $y_W$ at RHIC for $\sqrt{s} = 500$~GeV.  The different curves
follow the convention of Figure~1.}

\item[Figure 8]{The rapidity distribution for single $e^+$ and $e^-$
production at RHIC.  The different curves follow the convention of
Figure~1.  Figure~8a shows the rapidity distribution for
$\sqrt{s} = 200$~GeV and Figure~8b show the rapidity distribution for
$\sqrt{s} = 500$~GeV.}

\item[Figure 9]{As Fig.~4, for a) RHIC at 200~GeV and b) RHIC at 500~GeV.}

\item[Figure 10]{The charge asymmetry $A_e$ in single lepton production
(as defined in the text) versus $y_e$ at RHIC.  The different curves
follow the convention of Figure~1.  Figure~10a shows $A_e$ at
$\sqrt{s} = 200$~GeV and Figure~10b shows $A_e$ at $\sqrt{s} = 500$~GeV.}

\item[Figure 11]{The rapidity distribution for production of $W^-$ and
$W^+$ at RHIC, using EHLQ, set 2\cite{ehlq} parton distribution
functions.  The solid curve corresponds to $W^+$ production at
$\sqrt{s} = 500$~GeV, the dashed curve corresponds to $W^-$ production at
$\sqrt{s} = 500$~GeV, the dotted curve corresponds to $W^+$ production at
$\sqrt{s} = 200$~GeV and the dot-dashed curve corresponds to $W^-$
production at $\sqrt{s} = 200$~GeV.}

\item[Figure 12]{The asymmetry ${\cal A}_{LL}$ (as defined in the text)
versus $y_W$ at RHIC.  Figure~12a shows ${\cal A}_{LL}$ for
$\sqrt{s} = 200$~GeV and Figure~12b shows ${\cal A}_{LL}$ for
$\sqrt{s} = 500$~GeV.  We use EHLQ, set 2\cite{ehlq} for the unpolarized
parton distributions, while in both cases, the solid curve uses the
polarized parton distribution functions of BGC, set $I_S=0$ and the
dashed curve used the polarized parton distribution functions of BGC, set
$I_S=1$.}

\item[Figure 13]{The rapidity distribution for production of $e^-$ and
$e^+$ at RHIC, using EHLQ, set 2\cite{ehlq} parton distribution
functions.  The solid curve corresponds to $e^+$ production at
$\sqrt{s} = 500$~GeV, the dashed curve corresponds to $e^-$ production at
$\sqrt{s} = 500$~GeV, the dotted curve corresponds to $e^+$ production at
$\sqrt{s} = 200$~GeV and the dot-dashed curve corresponds to $e^-$
production at $\sqrt{s} = 200$~GeV.}

\item[Figure 14]{The asymmetry ${\cal A}^e_{LL}$ (as defined in the text)
versus $y_e$ at RHIC.  Figure~14a shows ${\cal A}^e_{LL}$ for
$\sqrt{s} = 200$~GeV and Figure~14b shows ${\cal A}^e_{LL}$ for
$\sqrt{s} = 500$~GeV.  We use EHLQ, set 2\cite{ehlq} for the unpolarized
parton distributions, while in both cases, the solid curve uses the
polarized parton distribution functions of BGC, set $I_S=0$ and the
dashed curve used the polarized parton distribution functions of BGC, set
$I_S=1$.}
\end{itemize}
\end{document}